\documentclass[twocolumn,prl,nofootinbib,floatfix,superscriptaddress,showpacs]{revtex4}

\usepackage{dcolumn}
\usepackage{graphicx}
\usepackage{rotating}
\usepackage{amsmath,amsfonts,amssymb}

\newcommand{\g}{\,\mathrm{g}}

\newcommand{\e}{\mathrm{e}}

\newcommand{\xmin}{\ensuremath{x_{\min}}}
\newcommand{\Mmin}{\ensuremath{M_{\min}}}
\newcommand{\weightn}{\ensuremath{\gamma_{n}}}

\newcommand{\remove}[1]{}

\begin{document}

\title{Evolutionary Model of Species Body Mass Diversification}
\author{A. Clauset}
\email{aaronc@santafe.edu}
\affiliation{Santa Fe Institute, 1399 Hyde Park Road, Santa Fe NM, 87501, USA}
\author{S. Redner}
\email{redner@bu.edu}
\affiliation{Center for Polymer Studies and Department of Physics, Boston University, Boston, MA 02215, USA}
\affiliation{Santa Fe Institute, 1399 Hyde Park Road, Santa Fe NM, 87501, USA}

\pacs{87.23.Kg, 02.50.-r}

\begin{abstract}
  
  We present a quantitative model for the biological evolution of species
  body masses within large groups of related species, e.g., terrestrial
  mammals, in which body mass $M$ evolves according to 
  branching (speciation), multiplicative diffusion, and an extinction
  probability that increases logarithmically with mass.  We describe this
  evolution in terms of a convection-diffusion-reaction equation for $\ln M$.
  The steady-state behavior is in good agreement with empirical data on
  recent terrestrial mammals, and the time-dependent behavior also agrees
  with data on extinct mammal species between $95$ -- $50$ Myr ago.

\end{abstract}

\maketitle

Animals---both extant and extinct---exhibit an enormously wide range of body
sizes.  Among extant terrestrial mammals, the largest is the African savannah
elephant ({\em Loxodonta africana africana}) with a mass of $10^{7}\g$, while
the smallest is Remy's pygmy shrew ({\em Suncus remyi}) at a diminutive
$1.8\g$. Yet the most probable mass is $40\g$, roughly the size of the
common Pacific rat ({\em Rattus exulans}), is only a little larger than the
smallest mass. More generally, empirical surveys suggest that such a broad
but asymmetric distribution in the number of species with adult body mass $M$
typifies many animal
classes~\cite{stanley:1973,kozlowski:gawelczyk:2002,allen:etal:2006,carrano:2006,meiri:2008},
including mammals, birds, fish, insects, lizards and possibly dinosaurs.

What mechanisms cause species mass distributions to assume such a shape?  A
satisfactory answer would have wide implications for the evolution and
distribution of the many other species characteristics that correlate with
body mass, including life span, metabolic rate, and extinction
risk~\cite{calder:1984,cardillo:etal:2005}.  Previous explanations for the
species mass distribution focused on detailed ecological, environmental and
species-interaction assumptions~\cite{allen:etal:2006}.  However, empirical
data present confusing and often inconsistent support for these theories, and
none explicitly address how species body mass distributions diversify in
time.

In this Letter, we construct a physics-based convection-diffusion-reaction
model to account for the evolution of the species mass distribution.  The
steady-state behavior of this model was recently solved to explain the
species mass distribution for recent terrestrial mammals and
birds~\cite{clauset:erwin:2008,clauset:etal:2009}, where recent is
conventionally defined as within the past 50,000
years~\cite{smith:etal:2003}.  Here we substantially extend this approach to
give predictions on mammalian body mass evolution that are in good agreement
with fossil data.  This model can further be used to estimate the historical
rates of body mass diversification from fossil data, which are otherwise
estimated using {\em ad hoc} techniques. To illustrate this application, we
estimate body mass diversification rates from our fossil data, which are in
good agreement with estimates of genetic diversification from molecular clock
methods~\cite{spring:etal:2003}.  Although our focus is on terrestrial mammal
evolution, this model can, in principle, be applied to any group of related
species.

Let $c(x,t)$ denote the number (density) of species having logarithmic mass
\mbox{$x=\ln M$} at a time $t$; we use $x$ as the basic variable in keeping
with widespread usage in the field \cite{clauset:erwin:2008}.
Our model incorporates three fundamental and empirically-supported features
of biological evolution.
\begin{enumerate}
\item Branching multiplicative
  diffusion~\cite{stanley:1973,clauset:erwin:2008}: each species of mass $M$
  produces descendant species (cladogenesis) with masses $\lambda M$, where
  $\lambda$ is a random variable, and the sign of the average
  $\langle\ln\lambda\rangle$ denotes bias toward larger or smaller
  descendants.  Empirical evidence~\cite{clauset:erwin:2008} suggests that
  $\langle\ln\lambda\rangle>0$ (known as Cope's rule \cite{stanley:1973}) for
  terrestrial mammals.
%
\item Species become extinct independently, and with a probability that
  increases weakly with mass~\cite{liow:etal:2008}.
\item No species can be smaller than a mass $M_{\min}$, due, for example, to
  metabolic constraints~\cite{west:etal:2002}.
\end{enumerate}

The production of descendant species corresponds to growth in the number of
species at a rate $k$ that is proportional to the density $c$ itself.
Similarly, the probability $p(x)$ that a species of logarithmic mass $x$
becomes extinct may also be represented by a loss term that is proportional
to $c$, but with a weak mass dependence~\cite{clauset:erwin:2008}.  We make
the simple choice of linear dependence $p(x)=A+Bx$ (but see below).  With
these ingredients, $c(x,t)$ obeys the convection-diffusion-reaction equation
in the continuum limit
\begin{equation}
\label{eq:cxt}
\frac{\partial c}{\partial t} + 
v \frac{\partial c}{\partial x} =
D \frac{\partial^2 c}{\partial x^2} + (k-A-Bx)c \enspace ,
\end{equation}
with bias velocity $v=\langle\ln\lambda\rangle$ and diffusion coefficient
$D=\langle (\ln\lambda)^2\rangle$, and where $k-A$ sets the absolute scale of species body mass frequencies.

To solve Eq.~\eqref{eq:cxt}, we substitute the eigenfunction expansion
$c(x,t)=\sum_{n} A_n \mathcal{C}_n(x) \, \e^{-\weightn t}$, which yields
\begin{equation}
\label{eq:cn-ode}
-\frac{\weightn}{D} \mathcal{C}_n + \mu \mathcal{C}_n' =  \mathcal{C}_n'' +(\alpha-\beta x) \mathcal{C}_n \enspace ,
\end{equation}
where the prime denotes differentiation with respect to $x$, $\mu=v/D$,
$\alpha=(k-A)/D$, and $\beta=B/D$. We eliminate the first derivative term by
introducing $\psi_n =\e^{-\mu x/2}\,\mathcal{C}_n$ and then we use the scaled variable
\mbox{$z=\beta^{1/3}
  x-\beta^{-2/3}\left(\alpha-\frac{\mu^2}{4}+\frac{\weightn}{D}\right)$} to
transform Eq.~\eqref{eq:cn-ode} into Airy's differential
equation, $\psi'' -z\psi=0$ \cite{abramowitz:stegun:1972},
%
%
for each eigenfunction $\psi_{n}$. 
The general solution is $\psi(z) = a_1\, {\rm Ai}(z) +a_2\, {\rm Bi}(z)$,
where Ai$(z)$ and Bi$(z)$ are the Airy functions; here the prime now denotes
differentiation with respect to $z$.  Since there can be no species with
infinite mass and Bi$(z)$ diverges as $z\to\infty$, we set $a_2=0$, while
$a_1$ is determined by the initial condition. (We could also incorporate a
fixed maximum species body mass $M_{\max}$, but the analysis is more
complicated without revealing additional insights.)

\begin{figure}[t]
\begin{center}
\includegraphics[scale=0.47]{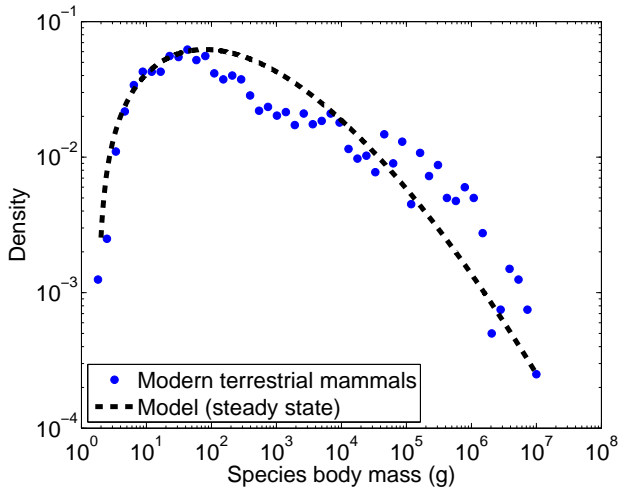} 
\end{center}
\caption{ (color online) Steady-state solution of the model for the species
  body mass distribution and suitably binned empirical data for 4002 recent
  terrestrial mammals (shown as a normalized histogram with 50
  logarithmically-spaced bins). }
\label{fig:steady:state}
\end{figure}

Since the species density vanishes at the minimum mass $\Mmin$, the argument of $\psi_{n}$ must equal 
\begin{align}
\label{eq:zn}
z_n  \!=\! \beta^{1/3}x_{\min} \!-\! \beta^{-2/3}\left(\!\alpha\!-\!\frac{\mu^2}{4}
\!+\!\frac{\weightn}{D}\!\right) 
  = z_0 \!-\!\frac{\weightn}{D\beta^{2/3}} \enspace .
\end{align}
The functions $\psi_{n}={\rm Ai}(z_{n})$ form the complete set of states for the eigenfunction expansion~\cite{courant:hilbert:1989}. 
The first few zeros $z_n$ are at (roughly) $-2.3381$, $-4.0879$, $-5.5205$
and $-6.7867$ for $n=0,1,2,3$~\cite{abramowitz:stegun:1972}.  (For computing
the distribution, we tabulated the first million zeros numerically using
standard mathematical software.)  The corresponding decay rates $\weightn$
are then given by $\weightn= D\,\beta^{2/3}(z_0-z_n)$, with $\gamma_{0}=0$ to give the steady state solution~\cite{clauset:etal:2009}.  These rates form an
increasing sequence so that the higher terms in the eigenfunction expansion
decay more quickly in time.  Finally, solving Eq.~\eqref{eq:zn} for $\alpha$ and plugging the result into 
the definition of $z$, we can eliminate the scale parameter $\alpha$ and
write \mbox{$z=z_n+\beta^{1/3}(x-\xmin)$}.
Thus each eigenfunction has the form
\begin{equation}
\label{eq:cn}
\mathcal{C}_n(x)  \propto \e^{\mu x/2}\, {\rm Ai}\left[ z_n+\beta^{1/3}(x-\xmin) \right],
\end{equation}
in which ${\rm Ai}(z)\sim \e^{-2z^{3/2}/3}$ for large $z$.  The competition
between this decay and the prefactor $\e^{\mu x/2}$ in $\mathcal{C}_n(x)$
contributes to the broadness of the species mass distribution and the
location of the most probable mass (Fig.~\ref{fig:steady:state}).
Parenthetically, the asymptotic decay of the eigenfunctions depends only weakly on the form of the extinction probability $p(x)$.  For instance, if we choose $p(x)=A+B x^{\delta}$, then as $z\to\infty$, the eigenfunctions decay as $\e^{-z^{1+\delta/2}}$.

\begin{figure}[t]
\begin{center}
\includegraphics[scale=0.47]{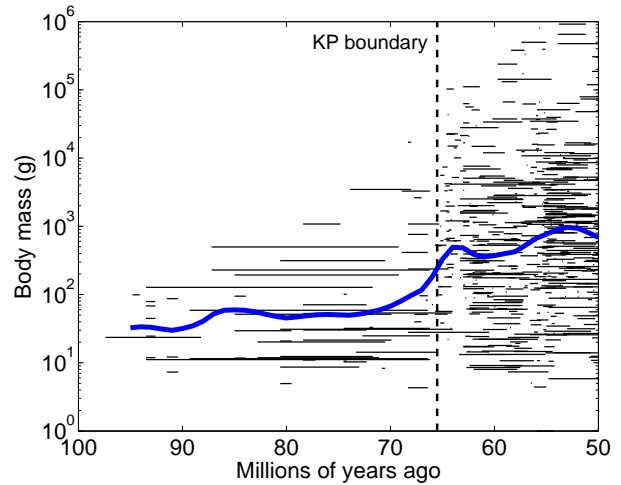}
\end{center}
\caption{(color online) Data on species body mass for 569 North American
  terrestrial mammals~\cite{alroy:2008} from 95 -- 50 Myr ago. Each horizontal
  segment represents a species, and endpoints denote its first and last
  appearance in the fossil record. The superimposed curve shows the average
  of $\ln M$ for these species (smoothed with an exponential kernel).
} 
\label{fig:fossil:data}
\end{figure}

Suppose that a given group of animals began its evolutionary history with a
single species of mass $M_0$ (initial condition $c(x,t=0) = \delta(x-x_0)$,
with $x_0=\ln M_0$), after which speciation occurs according to the dynamics
of Eq.~\eqref{eq:cxt}.  We use the fact that the Airy differential equation
is a Sturm-Liouville problem~\cite{courant:hilbert:1989} so that the
$\{\mathcal{C}_n\}$ form a complete and orthonormal set.  Following the
standard prescription to determine the coefficients of the eigenfunction
expansion, the full time-dependent solution is
\begin{equation}
\label{eq:c-soln}
c(x,t)= \sum_n \mathcal{C}_n(x_0)\mathcal{C}_n(x)\, \e^{-\weightn t} \enspace ,
\end{equation}
where each $\mathcal{C}_n(x)$ is given by Eq.~\eqref{eq:cn}.  In the long-time limit,
all the decaying eigenmodes with $n>0$ become negligible and the species
mass distribution reduces to $c(x,t\to\infty) \propto \mathcal{C}_0(x)$.

We test our predictions for the species body mass distribution by comparing
with available empirical data.  In the steady state, our model is
characterized by three parameters: $M_{\min}$, the mass of the smallest
animal, $\mu$, which controls the tendency of descendant species to be larger
or smaller than their ancestors, and $\beta$, which controls the dependence
of the extinction rate on mass.  The former two can be estimated directly
from  fossil data, while the latter is typically estimated by
matching the steady-state solution to the recent data.  For terrestrial
mammals, we previously found \mbox{$\mu\approx 0.2$}, $\beta\approx 0.08$,
while \mbox{$\Mmin\approx2\g$}~\cite{clauset:etal:2009}.  Using these parameter
values in the long-time limit, we obtain a good agreement between the
predictions of the model and the species mass distribution of recent
terrestrial mammals~\cite{smith:etal:2003} (Fig.~\ref{fig:steady:state}).

\begin{figure}[t]
\begin{center}
\mbox{\hskip -0.3in \includegraphics[scale=0.535]{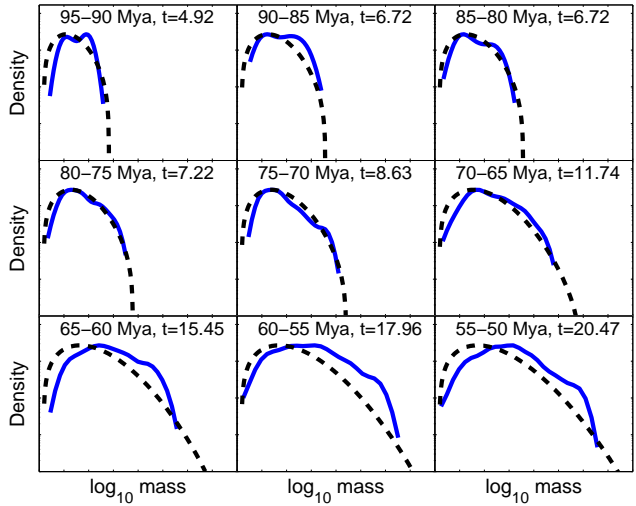}}
\end{center}
\vspace{-9.5mm}
\caption{(color online) A comparison of our model predictions (dashed) from
  Eq.~\eqref{eq:c-soln} and the mass distributions of extinct North American
  mammal species (solid) in nine consecutive time ranges. (To get reasonable results with sparse empirical data, we smooth the distributions with a Gaussian kernel.) }
\label{fig:snapshots}
\end{figure}

Our model also makes predictions about the way the body mass distribution
changes over time, which can be tested with fossil data.  Drawing on data
from the best available source on the evolution of mammalian body
masses~\cite{alroy:2008}, we plot in Fig.~\ref{fig:fossil:data} the durations
and body masses of 569 extinct mammal species from 95 -- 50 Myr ago.  This period includes the Cretaceous-Paleogene (KP)
boundary 
of 65.5 Myr ago that marks a mass extinction event during
which more than 50\% of then extant species became extinct, including
non-avian dinosaurs---the dominant fauna for the preceding 160 Myr---and is
the subject of many studies regarding the diversification of mammals. (The
Cretaceous period is conventionally abbreviated ``K,'' after the German
translation \textit{Kreide}.)

Using the same model parameters as above, and setting $M_{0}=2\g$, the
estimated size of the first mammal~\cite{luo:etal:2001},
Fig.~\ref{fig:snapshots} shows good agreement between model predictions from
Eq.~\eqref{eq:c-soln} and empirical data from Fig.~\ref{fig:fossil:data}. To
simplify the comparison, we divided the period from 95 -- 50 Myr ago into nine
bins of equal durations, and tabulated the distribution of species extant
during each of these bins.  In each of the nine panels of
Fig.~\ref{fig:snapshots}, we give both the historical time period and the
corresponding model time that yields a good qualitative match to the
data. (The fitting can be made more objective using standard techniques, but
the results are largely the same.)

\begin{figure}[t]
\begin{center}
\includegraphics[scale=0.47]{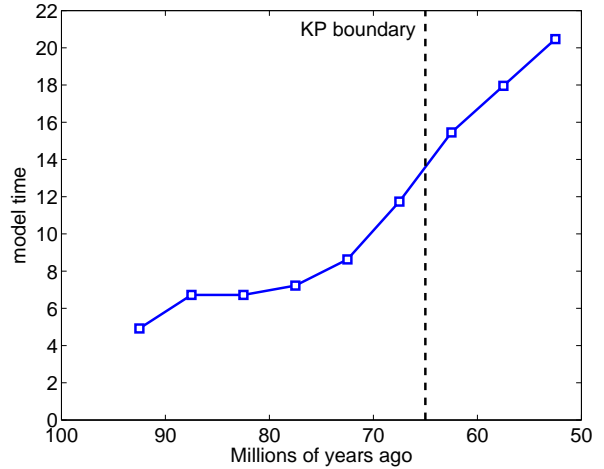} 
\end{center}
\caption{Estimated relation between model time to historical
  time, indicating that the broadening of the mammalian species mass
  distribution began approximately 10 Myr before the KP boundary when
  non-avian dinosaurs became extinct.  }
\label{fig:mapping}
\end{figure}

The relation between historical time and model time is itself interesting
(Fig.~\ref{fig:mapping}).  During the first 20 Myr (95 -- 75 Myr ago), the species
mass distribution is almost stationary, with model time advancing by only
$\Delta t=2.30$; curiously, between 90 -- 80 Myr ago, model time does not advance at all.  This period of near stasis may indicate a lull in the evolutionary dynamics, perhaps due to implicit competition from larger species, e.g., dinosaurs.  Over the next 20 Myr (75 -- 55 Myr ago), however, the distribution broadens considerably (model time advancing by $\Delta t = 10.74$), and comes to closely resemble the recent distribution (Fig.~\ref{fig:steady:state}). This correspondence suggests that the diversification of mammalian body masses into their recent state began at least 75 Myr ago, roughly 10 Myr before the KP boundary and the extinction of the non-avian dinosaurs. This estimate of the timing of body mass diversification for mammals agrees closely with some estimates of the timing of mammalian genetic diversification from studies of molecular clocks~\cite{spring:etal:2003}, and supports the notion that mammals were diversifying prior to and independently of the KP boundary itself~\cite{alroy:1999}. Whether these two forms of diversification are causally linked, however, is unknown.

Here we have held all model parameters constant while adjusting model time to fit the data. In principle, however, model time could advance steadily while varying some model parameters, perhaps to reflect large-scale changes or trends in the selection pressures on species body size. Empirical evidence supports a stable value of \Mmin~(Fig.~\ref{fig:fossil:data}), but little is currently known about how or why $\mu$ and $\beta$ may have varied.

As is typically the case with historical inference using the fossil record, 
a few caveats are in order.  Our fossil data are derived from the well-studied North American region using modern dental techniques, which are less prone to biases than older techniques. Still, some biases and sampling gaps likely persist, and these may explain the slight overabundance of large species, and underabundance of small species, in more recent times (Fig.~\ref{fig:snapshots}).
More significantly, recent fossil
discoveries suggest that, since mammals originated 195
Myr ago~\cite{luo:etal:2001}, mammalian diversification has proceeded in several
waves, and the vast majority of species groups from the earlier waves are now
extinct. Our data cover only the most recent diversification, in which
therian (placental and marsupial) mammals largely replaced the then dominant
non-therian mammal groups~\cite{luo:2007}. Unfortunately, suitable data on
these waves of diversification is not currently available, and the data we do
have is sparse in its coverage of non-therians. Thus, the application of our
model to infer diversification rates should be considered as a
proof-of-concept, illustrating that a physics-style model can shed
considerable light on evolutionary dynamics by placing the fossil record
within a theoretical framework.

In summary, the broad distribution of body masses for mammals appears to be
well described by a simple convection-diffusion-reaction model that
incorporates a small number of evolutionary features and constraints.
Indeed, our model does not account for many canonical ecological and
microevolutionary factors, such as interspecific competition, geography,
predation, and population dynamics~\cite{allen:etal:2006}. The fact that our
model agrees with species mass data suggests that the contributions of the
above-mentioned processes to the global character of body mass distributions
can be compactly summarized by the parameters $\mu$ and $\beta$ in our
model. Despite the crudeness of the model, the agreement between its
predictions and the available empirical data is satisfying.

Our model opens up intriguing directions for theoretical descriptions of
evolutionary dynamics.  For instance, our model ignores the populations of
individuals within each species; estimating the sizes of these populations
from body mass vs.\ population density scaling
relationships~\cite{white:etal:2007} and body mass vs.\ home-range size
relationships~\cite{brown:etal:1996:b} may allow us to calculate both the
total biomass contained in a group of related species, and its temporal
dynamics during diversification. Similarly, when paired with scaling
relations between body mass and metabolism~\cite{west:etal:2002}, we may be
able to calculate the total metabolic flux of a taxonomic group.

\begin{acknowledgments}
\noindent 
We thank D. H. Erwin, D. J. Schwab, Z.-X. Luo and J. Okie for helpful
comments, and J. Alroy, A. Boyer and F. Smith for kindly sharing data.  SR
gratefully acknowledges support from NSF grant DMR0535503.  This work was
also supported in part by the Santa Fe Institute.
\end{acknowledgments}


\end{document}